\documentclass[]{spie}  %>>> use for US letter paper
\usepackage[]{graphicx}
\usepackage{subcaption}
\usepackage{amsmath,amssymb,mathtools,url}
\title{\Large \bf The vortex fiber nulling mode of the Keck Planet Imager and Characterizer (KPIC)}
\usepackage{array}
\usepackage{hyperref}
\newcolumntype{P}[1]{>{\centering\arraybackslash}p{#1}}

\usepackage{xcolor}

\author{
Daniel Echeverri\supit{a,$\dagger$}, Garreth~Ruane\supit{b}, Nemanja~Jovanovic\supit{a}, Thomas~Hayama\supit{a}, Jacques-Robert~Delorme\supit{a}, Jacklyn~Pezzato\supit{a}, Charlotte~Bond\supit{c}, Jason~Wang\supit{a}, Dimitri~Mawet\supit{a,b} J.~Kent~Wallace\supit{b}, Eugene~Serabyn\supit{b} \skiplinehalf
\supit{a}Department of Astronomy, California Institute of Technology, 1200 E. California Blvd., \\Pasadena, CA 91125, USA \skiplinehalf
\supit{b}Jet Propulsion Laboratory, California Institute of Technology, 4800 Oak Grove Dr., \\Pasadena, CA 91109, USA \skiplinehalf
\supit{c}Institute for Astronomy, University of Hawaii, 640 North A'Ohoku Place, Hilo, HI 96720, USA
}

\authorinfo{$^\dagger$Send correspondence to dechever@caltech.edu}

\graphicspath{{Figures/}}
\begin{document} 
  \maketitle 

%%%%%%%%%%%%%%%%%%%%%%%%%%%%%%%%%%%%%%%%%%%%%%%%%%%%%%%%%%%%% 
\begin{abstract}
The Keck Planet Imager and Characterizer (KPIC) is an upgrade to the Keck II adaptive optics system that includes an active fiber injection unit (FIU) for efficiently routing light from exoplanets to NIRSPEC, a high-resolution spectrograph. Towards the end of 2019, we will add a suite of new coronagraph modes as well as a high-order deformable mirror. One of these modes, operating in $K$-band (2.2$\mu m$), will be the first vortex fiber nuller to go on sky. Vortex Fiber Nulling (VFN) is a new interferometric method for suppressing starlight in order to spectroscopically characterize exoplanets at angular separations that are inaccessible with conventional coronagraph systems. A monochromatic starlight suppression of $6\times10^{-5}$ in 635~nm laser light has already been demonstrated on a VFN testbed in the lab. A polychromatic experiment is now underway and coupling efficiencies of $<5\times10^{-4}$ and $\sim5\%$ have been demonstrated for the star and planet respectively in 10\% bandwidth light. Here we describe those experiments, the new KPIC VFN mode, and the expected performance of this mode using realistic parameters determined from on-sky tests done during the KPIC commissioning. 
\end{abstract}

%>>>> Include a list of keywords after the abstract 

\keywords{Instrumentation, exoplanets, nulling interferometry}

%%%%%%%%%%%%%%%%%%%%%%%%%%%%%%%%%%%%%%%%%%%%%%%%%%%%%%%%%%%%%
\section{INTRODUCTION}
\label{sec:intro} 

The Keck Planet Imager and Characterizer (KPIC)\cite{Mawet2016_KPIC,Mawet2017_KPIC,Jovanovic2019SPIE} is an instrument designed to link the Keck adaptive optics (AO) system\cite{Wizinowich2000} to NIRSPEC\cite{Martin2018}, a high-resolution near-infrared spectrograph. KPIC accomplishes this by coupling light from a point-like source into a single-mode fiber (SMF)\cite{Jovanovic2017}, allowing for conventional stellar spectroscopy as well as direct spectroscopy of low mass companions, including giant exoplanets. In scenarios where the companion of interest is resolved with respect to the star (roughly speaking, when the angular separation is $>\lambda/D$, where $\lambda$ is the wavelength and $D$ is the telescope diameter), the starlight may be minimized at the position of the planet using a coronagraph and speckle nulling techniques\cite{Pezzato2019SPIE,Mawet2017_HDCII}. However, in situations where the planet-star angular separation is $\sim \lambda/D$, fiber nulling interferometry\cite{Bracewell1978,Haguenauer2006,Serabyn2019PFN} may be a more effective method for reducing the amount of starlight entering the spectrograph. KPIC will have such an interferometric mode, based on the vortex fiber nulling (VFN) technique\cite{Ruane2018_VFN,Echeverri2019_VFN,Ruane2019SPIE}, to enable spectroscopy of close-in companions. In this paper, we provide an overview of the VFN concept, an update on the laboratory demonstrations of VFN, and estimates of the on-sky performance of the KPIC VFN mode given the current AO performance. 

\section{VFN Concept}
\label{sec:concept}
VFN is an interferometric method for detecting and spectroscopically characterizing exoplanets at small angular separations that are inaccessible to conventional coronagraph instruments\cite{Ruane2018_VFN}. It leverages the modal selectivity of SMFs to reject starlight while efficiently coupling planet light that can then be fed into a spectrograph for analysis.

\begin{figure}[t!]
    \centering
    \includegraphics[width=\linewidth]{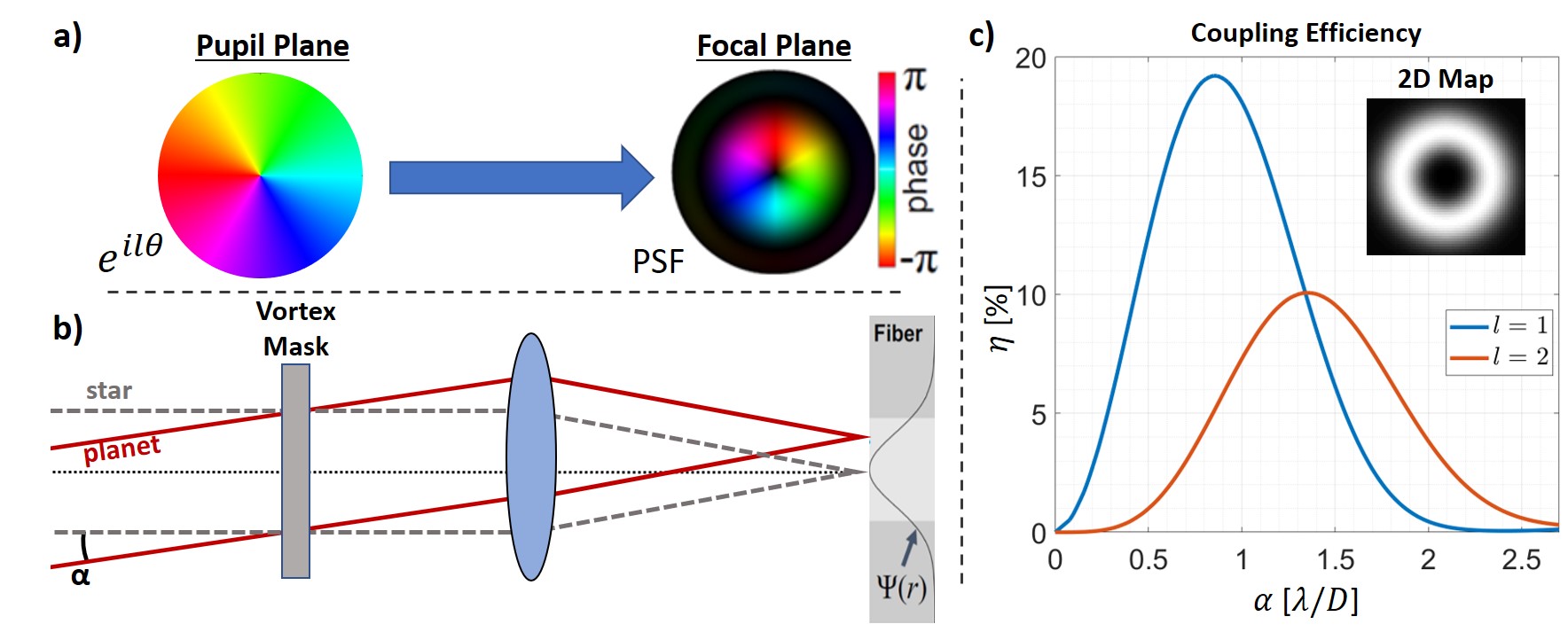}
    \caption{(a) The azimuthally varying phase pattern introduced by a charge $l=1$ vortex mask and the corresponding complex-valued PSF resulting when the vortex mask is placed in a pupil plane. (b) Diagram of a VFN system with the vortex mask in the pupil plane and SMF in the image plane. The mode of the fiber is denoted $\Psi(r)$. (c) Coupling efficiency, $\eta$, of a point source versus its angular separation from the optical axis, $\alpha$, for a charge $l=1$ (blue) and $l=2$ (orange) VFN system. The inset shows the coupling efficiency for all points in a field of view centered on the star/fiber.}
    \label{fig:MotherFig}
\end{figure}

A vortex phase mask placed in the pupil plane\cite{Swartzlander2001} introduces an azimuthally increasing phase ramp pattern, shown in the left half of Fig.~\ref{fig:MotherFig}a, of the form $\exp(il\theta)$, where $\theta$ is the azimuthal coordinate and $l$ is an integer known as the charge which defines how many times the phase cycles around the beam. In theory, a VFN system can have the vortex mask located anywhere in the beam path. In the KPIC VFN mode, the vortex mask will be located in a pupil plane (see Fig.~\ref{fig:MotherFig}b). This results in the complex-valued point spread function (PSF) shown in the right half of Fig.~\ref{fig:MotherFig}a which is characterized by a ``donut shape" in amplitude and a vortex phase structure that also varies as $\exp(il\theta)$. 

The coupling efficiency, or fraction of light from a given source that gets into the SMF, can be computed as:
\begin{equation}
    \eta(\alpha)=\frac{\left|\int E(\mathbf{r};\alpha) \Psi(r)dA\right|^2}{\int \left| E(\mathbf{r};\alpha)\right|^2dA \int \left|\Psi(r)\right|^2dA} , %Normalized version
    %\eta(\alpha)=\left|\int E(\mathbf{r};\alpha) \Psi(r)dA\right|^2,
    \label{eqn:couplingeff}
\end{equation}
where $E(\mathbf{r};\alpha)$ is the field at the entrance to the SMF, $\Psi(r)$ is the fundamental mode of the SMF, $\mathbf{r}=(r,\theta)$ are the coordinates in the fiber-tip plane, and $\alpha$ is the angular offset with respect to the optical axis (see Fig.~\ref{fig:MotherFig}b). To give a heuristic explanation of why the starlight is rejected by the fiber, we consider a simplified case where the system pupil is circular and unobscured and the post-vortex stellar field is composed of separate radial and azimuthal components, $E_s(\mathbf{r})=f_r(r)\exp(il\theta)$. Since the fiber mode is also radially symmetric, the coupling integral for the star is separable and the azimuthal term goes to zero for any $l\neq 0$. Thus, the complex stellar field is orthogonal to $\Psi(r)$ and does not couple into the fiber. An off-axis point source has the same PSF but is shifted with respect to the SMF mode such that the azimuthal term no longer vanishes and the field is not orthogonal to $\Psi(r)$. Thus, the off-axis planet light partially couples. For non-circular pupils, such as the Keck aperture, the computation is different but a similar orthogonality condition occurs. 

Figure~\ref{fig:MotherFig}c shows the coupling efficiency of a point source as a function of the angular separation, $\alpha$, for the system shown in Fig.~\ref{fig:MotherFig}b. As expected, a complete rejection, or nulling condition, is obtained for the star since it is an on-axis point source, $\eta_s(\alpha=0)=0$. Meanwhile a peak coupling of 19\% is achieved for an off-axis point source, such as a planet, at an angular separation of $\alpha=0.85\lambda/D$ with a charge $l=1$ vortex. Similarly, a peak coupling of 10\% is obtained for $\alpha=1.3\lambda/D$ with a charge $l=2$ vortex. Following the conventional definition of inner-working-angle (IWA) as the separation where the planet throughput is $50\%$ of its peak value, a charge 1 VFN system has inner and outer-working angles of $0.4$ and $1.4\lambda/D$ respectively. The entire charge 1 VFN working region is thus within the typical IWA\cite{Guyon_2006} of conventional coronagraphs. Furthermore, the rotational symmetry of the coupling efficiency about the star allows for planets to be spectrally characterized even when the planet's position angle as projected on sky is unknown, which is often the case for planets detected through radial velocity (RV) and transit techniques. 

Ruane et al.\cite{Ruane2019SPIE} provide a more detailed explanation of the VFN concept, its benefits, and practical design considerations. Most importantly, VFN works well with Keck's segmented aperture and requires few modifications to the optical design of KPIC.

\section{Laboratory Demonstrations}
\subsection{Monochromatic Demonstration}
\label{sec:monochromatic}

\begin{figure}[t!]
    \centering
    \includegraphics[width=0.85\linewidth]{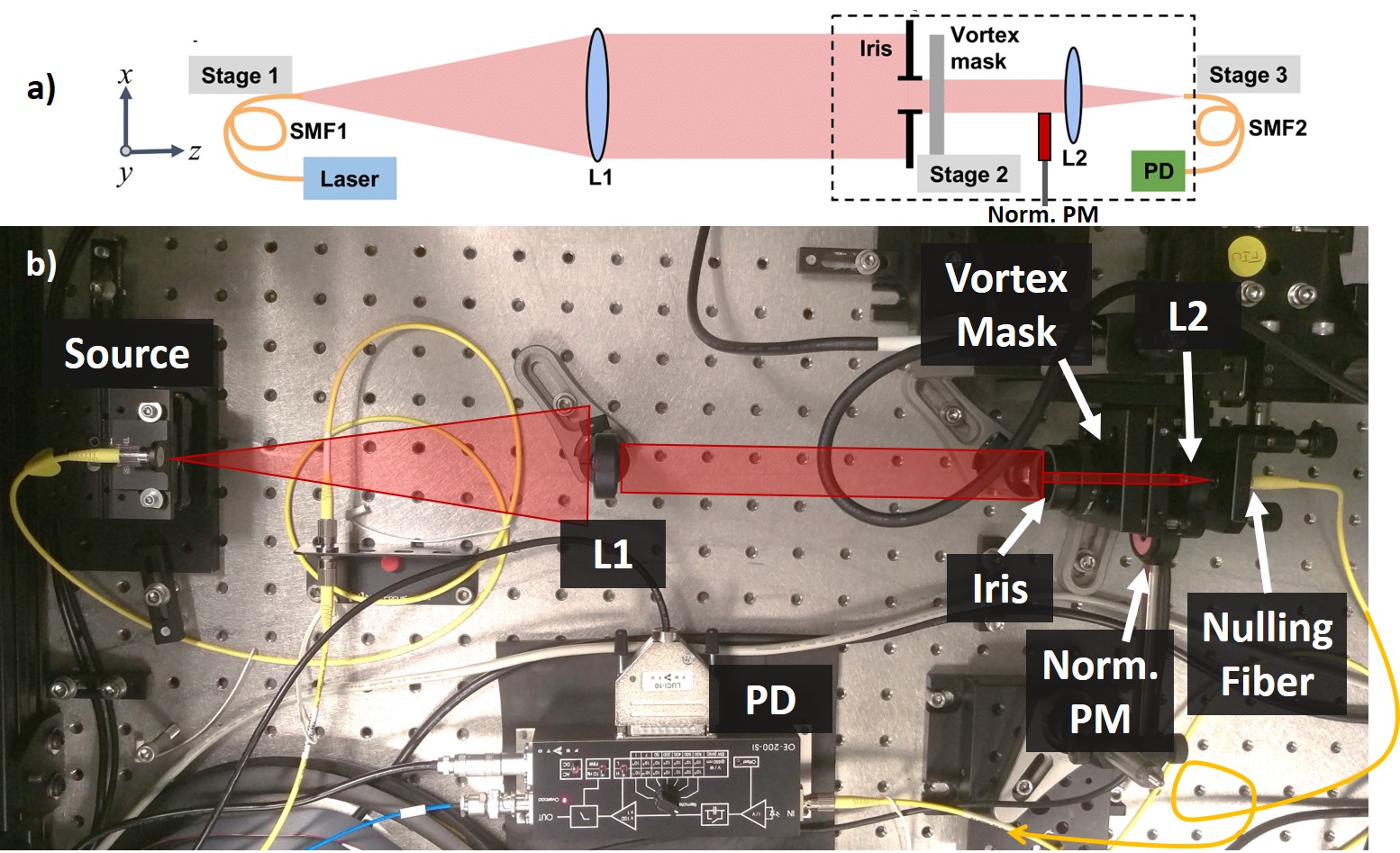}
    \caption{(a) Schematic of the transmissive VFN testbed at Caltech. Stage 1 holds the fiber source which projects light onto L1, the collimating lens. An iris then sets the pupil diameter before passing the beam to the vortex mask in stage 2. A focusing lens images the beam onto stage 3 which holds the nulling fiber that is connected to a photodiode for coupling measurements. A retractable power meter, norm. PM, can be moved into the beam path to measure the power for normalization. (b) Picture of the Caltech VFN testbed.}
    \label{fig:LabSetup}
\end{figure}

\begin{figure}[t!]
    \centering
    \includegraphics[width=\linewidth]{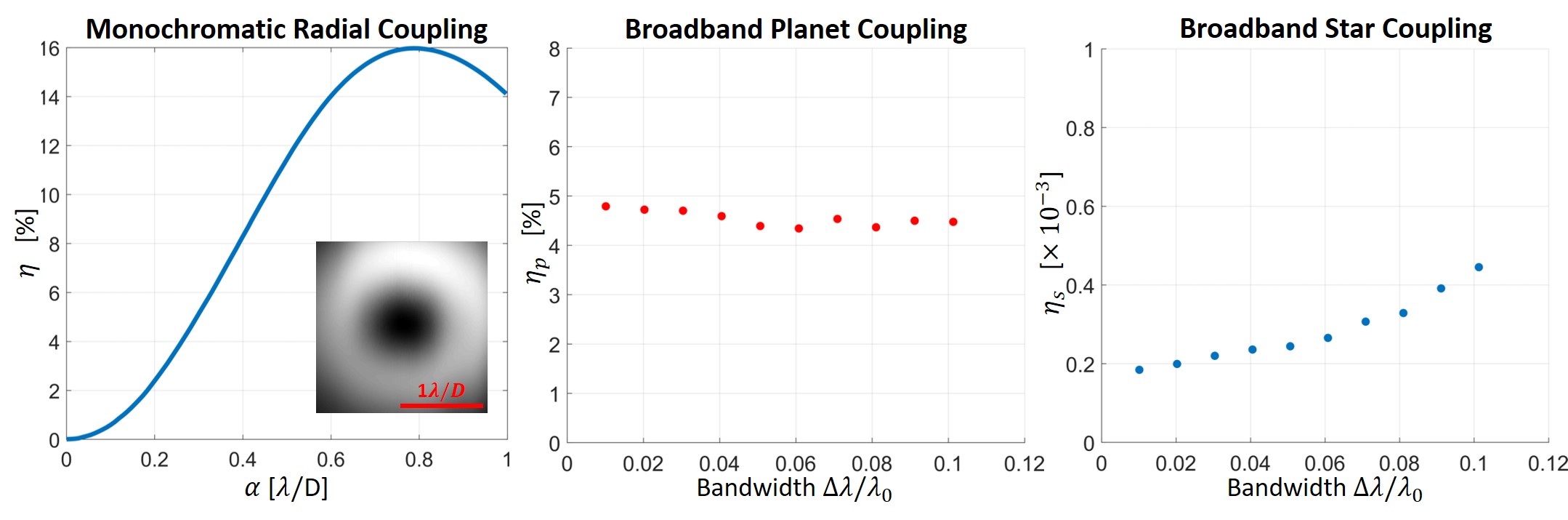}
    \caption{(left) Azimuthally averaged radial coupling, $\eta$, with 635~nm laser light and a charge 1, monochromatic vortex mask. The inset is the two dimensional coupling map. (middle) Azimuthally averaged peak planet coupling, $\eta_p$, in broadband light at various bandwidths centered around $\lambda_0=790$~nm and with a charge 2 polychromatic vortex mask. (right) Same as middle, but showing the null depth or ``stellar coupling", $\eta_s$.}
    \label{fig:LabResults}
\end{figure}

Earlier this year, the VFN concept was demonstrated in monochromatic light for the first time in the lab\cite{Echeverri2019_VFN}. We were able to achieve stellar coupling fractions, or ``null depths", of $6\times10^{-5}$ and an average peak planet coupling of $10\%$ using a monochromatic charge 1 vortex mask, a 635~nm laser source, and other commercial off-the-shelf optics (see Fig.~\ref{fig:LabSetup}). Furthermore, we predicted that the main reason for the reduced planet coupling relative to the theoretical $19\%$ for a charge 1 VFN system was an incorrect $F/\#$ in our setup. Thus, by correcting the $F\#$ and making a few minor modifications, we have since achieved an azimuthally averaged peak planet coupling of $16\%$ while still maintaining the same null depth (left plot of Fig.~\ref{fig:LabResults}). The inset 2D coupling map shows that the ``donut" is rounder and more symmetric than before. 

We believe that the remaining $3\%$ missing from the expected planet coupling is primarily due to: a) throughput losses in the final focusing lens and fiber and b) errors in the calibration between our two power meters. In our experiments, we measure the normalization power (denominator in Eq.~\ref{eqn:couplingeff}) just before the final lens using a retractable power meter, norm. PM, as shown in Fig.~\ref{fig:LabSetup}. We assume that the throughput losses in the lens and fiber match the manufacturers' specifications but we have not measured these losses precisely to confirm that they are correct. As such, this may provide a marginal deviation in coupling efficiency. The larger deviation could arise from the calibration between our two power meters. The normalization power is measured with a Thorlabs~S120C detector head but the coupled power (numerator in Eq.~\ref{eqn:couplingeff}) is measured on a separate fiber-coupled photodiode, \mbox{Femto~OE-200-SI-FC}, at the output of the nulling SMF. We carefully calibrated these two detectors relative to each other but there is some uncertainty in the accuracy of the calibration across the multiple gain settings needed for these measurements. As such, we believe that a more accurate coupling fraction can be obtained by modifying the system so that both the normalization and the coupled power can be measured on the same detector as explained in Sec~\ref{sec:testbedRedesign}.

\subsection{Polychromatic Demonstration}
\label{sec:polychromatic}

With some minor modifications, we were able to use the same testbed shown in Fig.~\ref{fig:LabSetup} to demonstrate the VFN concept in polychromatic light. We replaced the 635~nm laser with a supercontinuum white light source (NKT Photonics SuperK EXTREME) and tunable filter (NKT Photonics SuperK VARIA) which allows us to select the bandwidth and central wavelength within the visible spectrum. We also replaced the monochromatic charge 1 vector vortex mask with an analogous polychromatic charge 2 mask, manufactured by Beam Co.\cite{Tabiryan2017}, that nominally operates from $420\text{-}870$~nm. We found the wavelength that produced the best null with this mask by scanning through central wavelengths with 3~nm bandwidth and tuning the vortex and fiber positions.

The optimal wavelength was 790~nm, where we achieved a null depth of $2\times10^{-4}$ and an azimuthally averaged peak planet coupling of $\sim5\%$ with a 3~nm bandwidth. We then gradually increased the bandwidth without changing anything else in the system and measured the broadband coupling efficiencies shown in Fig.~\ref{fig:LabResults}. At 10\% bandwidth (defined as $\Delta\lambda/\lambda_0$), we achieved a broadband null of $4.2\times10^{-4}$ and an azimuthally averaged peak planet coupling of $\sim4.5\%$. Thus, there was very little degradation in performance with increasing bandwidth. The main limitation on the maximum bandwidth was the upper wavelength limit of 840~nm from the VARIA filter. With a longer-wavelength source, we would likely achieve larger bandwidths with similar performance. Furthermore, the slight decrease in peak planet coupling at the larger bandwidths is likely due to the fact that our transmissive system uses a singlet focusing lens with an inherently wavelength-dependent focal length. As such, the planet coupling at the edges of the band is reduced by defocus.

The peak planet coupling of 5\% is half of the ideal theoretical coupling for a charge 2 vortex. These coupling losses may be due to a few factors including that: a) we are using SM600 fibers at wavelengths beyond their specified range, b) the anti-reflection (AR) coating on the final lens, L2, is sub-optimal at wavelengths beyond 700~nm, and c) the pupil diameter is currently set to produce the ideal $F/\#$ for 635~nm light, not 790~nm. Even with this sub-optimal configuration, we believe this experiment demonstrates that VFN has the potential to provide broadband nulling over a typical astronomical band ($\sim$20\%). 

\subsection{Planned Testbed Upgrades}
\label{sec:testbedRedesign}

To improve the broadband performance and capabilities of the VFN testbed at Caltech, we have redesigned the system and are planning to implement the following upgrades in coming months:
\vspace{-4mm}
\begin{enumerate}
    \setlength\itemsep{-5pt}
    \item Replace the transmissive lenses with reflective off-axis parabolas (OAPs).
    \item Modify the source stage to use a pinhole and add space for polarizers.
    \item Replace the current fiber positioning actuators with larger-range, higher-precision actuators.
    \item Modify the fiber injection stage to support two fibers: the nulling SMF and a multi-mode fiber.
    \item Use a fiber coupler to combine the two fibers.
    \item Use anti-reflection (AR) coated fibers. 
\end{enumerate}
\vspace{-3mm}
The use of OAPs will minimize chromatic aberrations and especially remove the chromatic focal shift such that the planet and star coupling efficiencies will no longer be affected at the edges of the band when working with large bandwidths. This will also allow us to switch easily between central wavelengths which will be helpful as we transition from our current visible experiments to planned $K$-band $(2.2~\mu m)$ tests in the future. Using a pinhole ensures that the source remains unresolved which is particularly useful in the charge 1 VFN case since it is much more sensitive to the angular size of the source\cite{Ruane2018_VFN}. Furthermore, the new design creates space before the pinhole for polarizers to be placed without reducing the wavefront quality of the system. Filtering for a single polarization state will allow us to further investigate polarization-dependent aberrations in the vortex mask and reduce stellar leakage due to imperfect retardance.

The new fiber positioning actuators (PI~Q-545.240) will provide closed loop accuracy down to $\sim6$~nm. This will allow us to measure the coupling versus position more precisely while ensuring that we have the accuracy needed to center the fiber on the optimal null location. These actuators also have a 25~mm range which will allow us to easily switch between two fibers in the focal plane: the nulling SMF and a 400~micron core multi-mode fiber (MMF). The MMF will couple $>99.9\%$ of the light incident on the fiber plane for normalization purposes. Additionally, by using fibers with the same AR coating, we'll implicitly account for Fresnel reflections in the normalization as well. Using a fiber coupler to combine the MMF and SMF allows us to use the same photodiode for the coupling and normalization measurements which removes the need to cross calibrate two different power meters. These upgrades combined will provide more accurate coupling efficiency, $\eta$, measurements. 

\section{VFN with KPIC}
\label{sec:KPIC}

\subsection{The KPIC VFN Mode}
\label{KPICVFNMode}
\begin{figure}[t!]
    \centering
    \includegraphics[width=\linewidth]{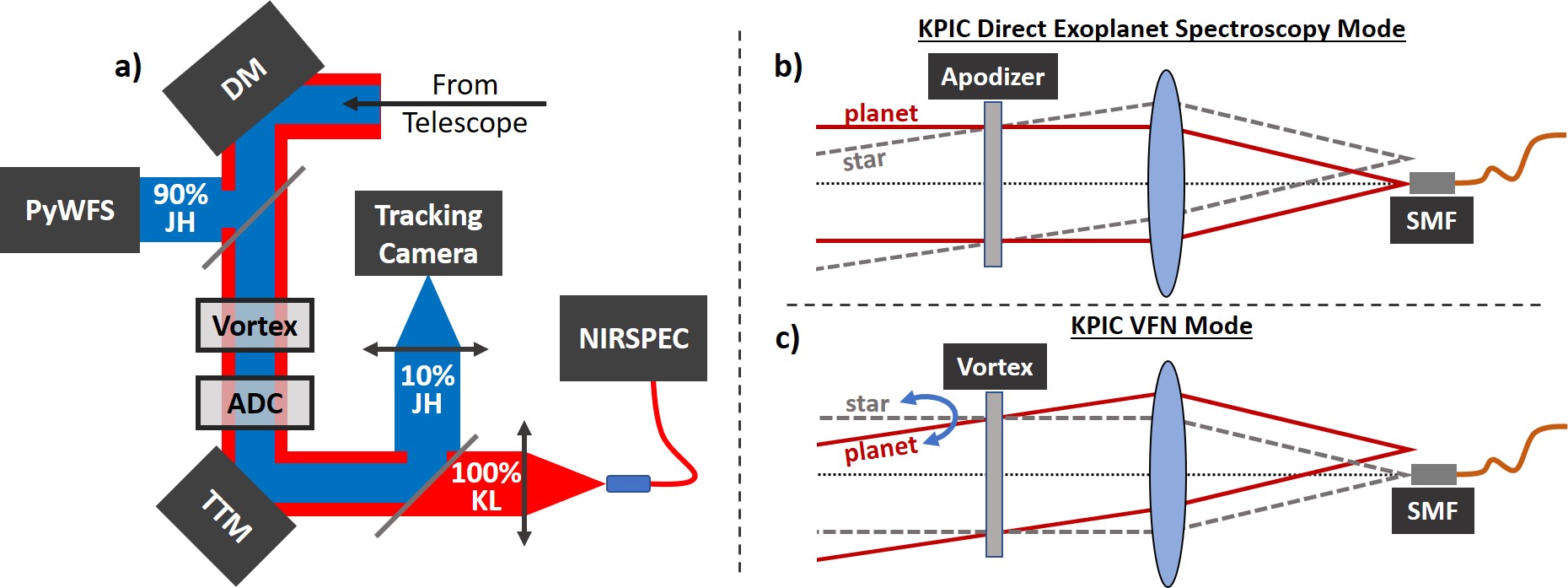}
    \caption{(a) Schematic of the KPIC VFN mode. Light arrives from the telescope after passing through the facility AO system. The near-infrared pyramid wavefront sensor (PyWFS) and high-order deformable mirror (DM) further correct the wavefront before allowing the beam to pass through the vortex mask and then the atmospheric dispersion compensator (ADC). The tip-tilt mirror (TTM) centers the star PSF on the fiber which feeds NIRSPEC, the high-resolution near-infrared spectrograph. A dichroic reflects $J$- and $H$-bands to a tracking camera, which provides simultaneous imaging for PSF tracking, calibration, and control algorithms. Light in the science channel ($K$-band) transmits through the dichroic and is routed to NIRSPEC via the SMF. $L$-band is unused in the initial VFN configuration. (b)~Nominal KPIC observation mode for direct exoplanet spectroscopy with the apodizer in the pupil and the planet aligned to the fiber. (c)~KPIC VFN mode with the vortex mask in the pupil and the star aligned to the fiber.}
    \label{fig:KPIC_VFNMode}
\end{figure}

Given the simplicity of this approach and the recent progress on laboratory demonstrations, we are preparing to add a VFN mode to the KPIC instrument as part of an upcoming upgrade. This VFN mode shares many modules with the other KPIC observation modes but there are a few key differences. As shown in Fig.~\ref{fig:KPIC_VFNMode}a, the VFN mode uses the near-infrared pyramid wavefront sensor\cite{Bond2018_PyWFS} (PyWFS) and high-order deformable mirror \mbox{(DM; Boston Micromachines kilo-DM)} to perform an additional wavefront control loop after the facility AO system. The corrected beam then passes through the vortex mask and atmospheric dispersion compensator (ADC). A tip-tilt mirror (TTM) aligns the stellar PSF with the science SMF that feeds NIRSPEC (see Fig.~\ref{fig:KPIC_VFNMode}c) to null the starlight. Feedback for the TTM control loop is provided primarily by the tracking camera which images the PSF just before the final focusing optics, but further feedback can also be obtained at a slower cadence using the slit-viewing camera of NIRSPEC. 

Thus, the VFN mode (Fig.~\ref{fig:KPIC_VFNMode}c) is slightly different from the direct exoplanet spectroscopy mode of KPIC (Fig.~\ref{fig:KPIC_VFNMode}b), which a) uses an optional apodizer instead of a vortex and b) aligns the planet with the fiber. The direct spectroscopy mode is better suited for characterization of known exoplanets at larger separations from their host-stars ($>\lambda/D$), while the VFN mode is better for blind or targeted surveys and characterization of close-in companions ($\sim\lambda/D$).

\begin{figure}[t!]
    \centering
    \includegraphics[width=\linewidth]{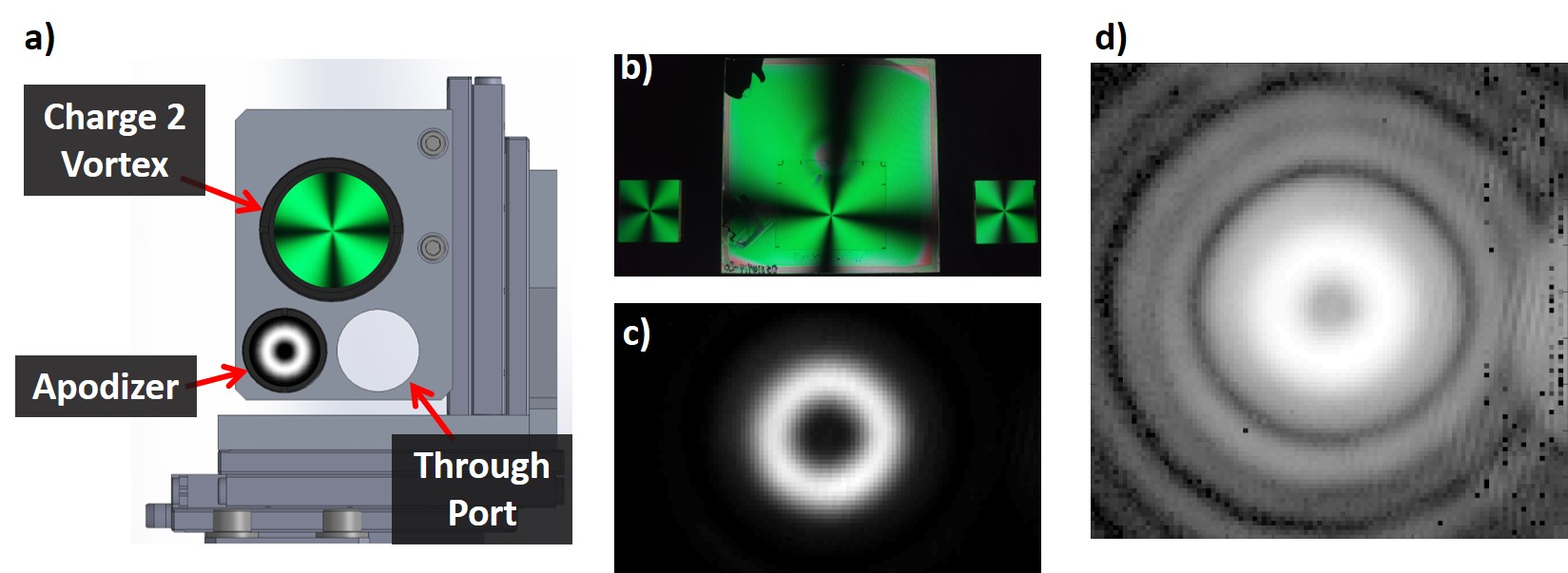}
    \caption{(a) Model of the pupil mask stage for KPIC. (b) Charge 2 $K$-band vector vortex masks being considered for deployment in KPIC as viewed in cross-polarization. (c) The PSF produced by the vortex masks in (b) when placed in a pupil plane, which takes on the expected donut shape. (d) Same as (c), but on a logarithmic scale.}
    \label{fig:KPIC_CoronagraphModule}
\end{figure}

In order to enable the VFN mode, a charge 2 $K$-band vector vortex mask will be installed in the pupil mask stage alongside the apodizer, as shown in Fig.~\ref{fig:KPIC_CoronagraphModule}a. Pezzato et al.\cite{Pezzato2019SPIE} describe this module and the custom-designed apodization mask it carries in detail. A charge 2 vortex mask was chosen for VFN based on the predicted and early on-sky performance of KPIC. Although a charge 1 vortex yields higher planet throughput at smaller angular separations which can significantly decrease the integration time needed to observe an exoplanet, this improved planet sensitivity comes at the cost of increased sensitivity to tip/tilt errors. For example, to achieve a null depth of $\eta_{s}=10^{-4}$, a charge 1 VFN system requires less than $0.01\lambda/D$ RMS tip-tilt jitter whereas a charge 2 needs $0.1\lambda/D$ for a similar null depth\cite{Ruane2019SPIE}. While a charge 2 vortex takes a hit in planet coupling, it relaxes tip/tilt requirements, which will be useful during the early stages of VFN development. As more on-sky measurements of the AO performance are made, we will reconsider whether a charge 1 or 2 vortex is optimal given the current system performance. 

We chose to start with $K$-band $(2.2\mu m)$ operation first for similar reasons; the wavefront errors scale with wavelength. Nevertheless, we have considered the possibility of including $H$- $(1.65\mu m)$ or $J$- $(1.25\mu m)$ band operation in the future and have left a clear path to implementing this capability if the AO performance allows for it. Longer wavelengths are also possible, but the performance may be limited by thermal background. 

Given these design considerations, we have started testing charge 2 $K$-band vortex masks in the lab (see Fig.~\ref{fig:KPIC_CoronagraphModule}b). We measured the transmission of these masks at $2\mu m$ to be $>99\%$. We also put these vortex masks in the pupil plane of a simple optical system to image their PSF. The resulting PSFs (Fig.~\ref{fig:KPIC_CoronagraphModule}c-d) show the expected donut pattern. We plan to further validate these masks with polychromatic coupling measurements on the upgraded VFN testbed as well as on a dedicated KPIC testbed at Caltech\cite{Pezzato2019SPIE}.

\subsection{Predicted On-Sky VFN Performance}
\label{sec:VFNSimulations}
To predict the performance of the KPIC VFN mode, we have started simulating the system assuming wavefront errors based on PyWFS measurements made during KPIC on-sky engineering runs. The PyWFS provides the residual wavefront error sampled at 1~kHz. Although the PyWFS beam path has optics that are non-common with the VFN beam path, we assume that the measurements represent the wavefront just before the vortex mask. In practice, image sharpening routines will be run on the tracking camera during the daytime before any KPIC observation run to minimize the non-common path aberrations. The PyWFS also provides the residual tip/tilt error at 1~kHz which we can feed into the simulator as well. For now, these simulations assume that the TTM is not being used for fast control so we apply the tip/tilt residuals from the PyWFS directly. Additionally, the simulator also accounts for the predicted residual atmospheric dispersion left over by the ADC.

\begin{figure}[t!]
    \centering
    \includegraphics[width=0.9\linewidth]{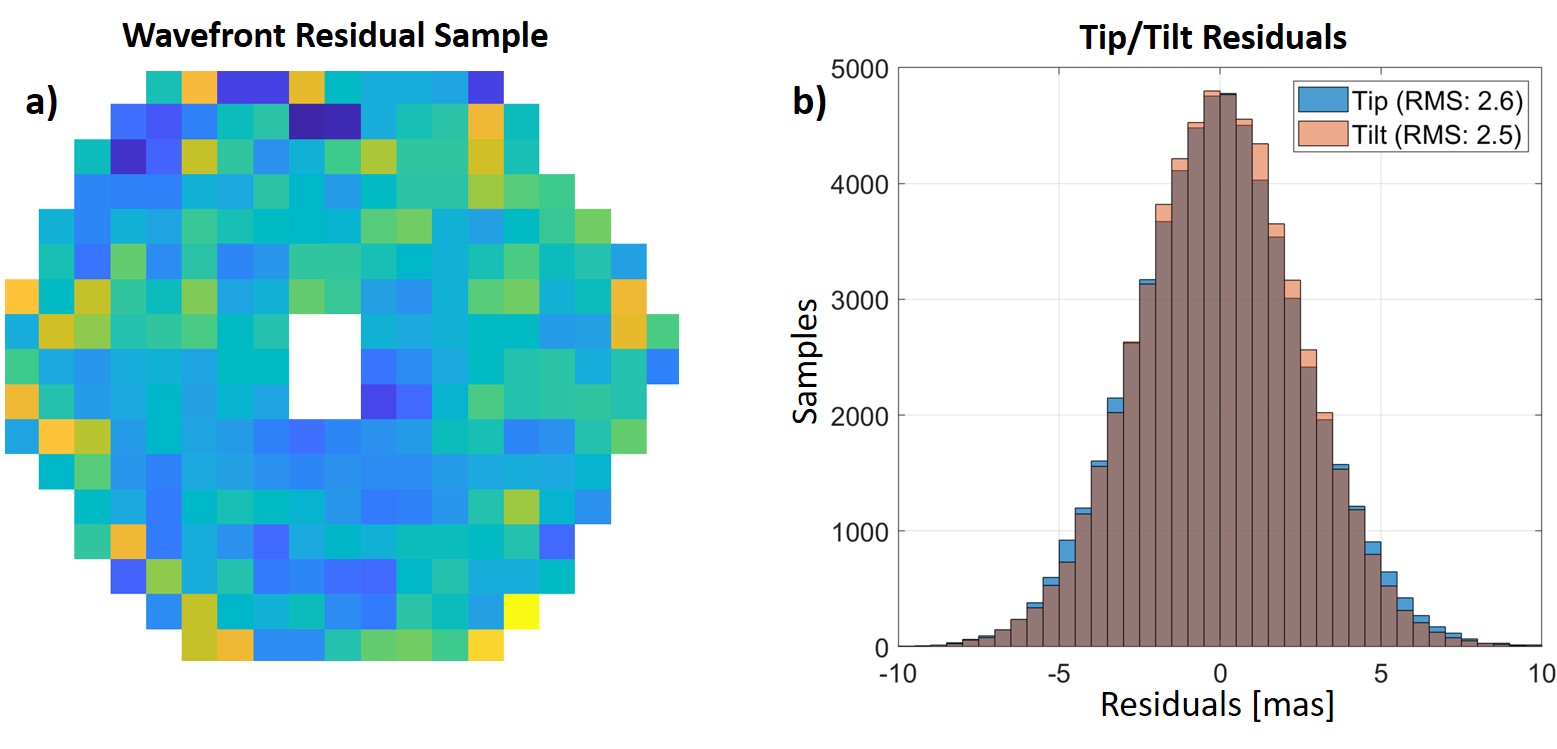}
    \caption{(a) Sample of the PyWFS wavefront residuals shown as projected onto the low-order facility DM. The average RMS wavefront residual over the 60 seconds of data was 150~nm. (b) Tip/Tilt residuals as reported by the PyWFS. The RMS tip and tilt residuals for the 1 minute sample are 2.6 and 2.5 mas respectively.}
    \label{fig:RawPyWFSData}
\end{figure}

For this paper, we used PyWFS data obtained on June 17\textsuperscript{th}, 2019 during an on-sky engineering run. Figure~\ref{fig:RawPyWFSData} shows the residual wavefront errors measured by the PyWFS. Figure~\ref{fig:RawPyWFSData}a is a sample of the wavefront residuals while Fig.~\ref{fig:RawPyWFSData}b has the tip/tilt residuals in milliarcseconds (mas) for the full 60 seconds of data at 1~kHz. The average seeing that night was about 0.6 arcseconds. The average RMS wavefront residuals in the minute-long sample were 150~nm while the tip/tilt residuals were 2.6 and 2.5 mas RMS respectively. The spatial wavefront sampling shown in Fig.~\ref{fig:RawPyWFSData}a is sufficient for simulating the VFN performance since the VFN is fairly insensitive to high-frequency aberrations\cite{Ruane2018_VFN}.

\begin{figure}[h!]
\begin{subfigure}{\textwidth}
    \centering
    \caption{Charge 2 VFN case as currently planned for KPIC (Online Video 1).}
    \includegraphics[width=\linewidth]{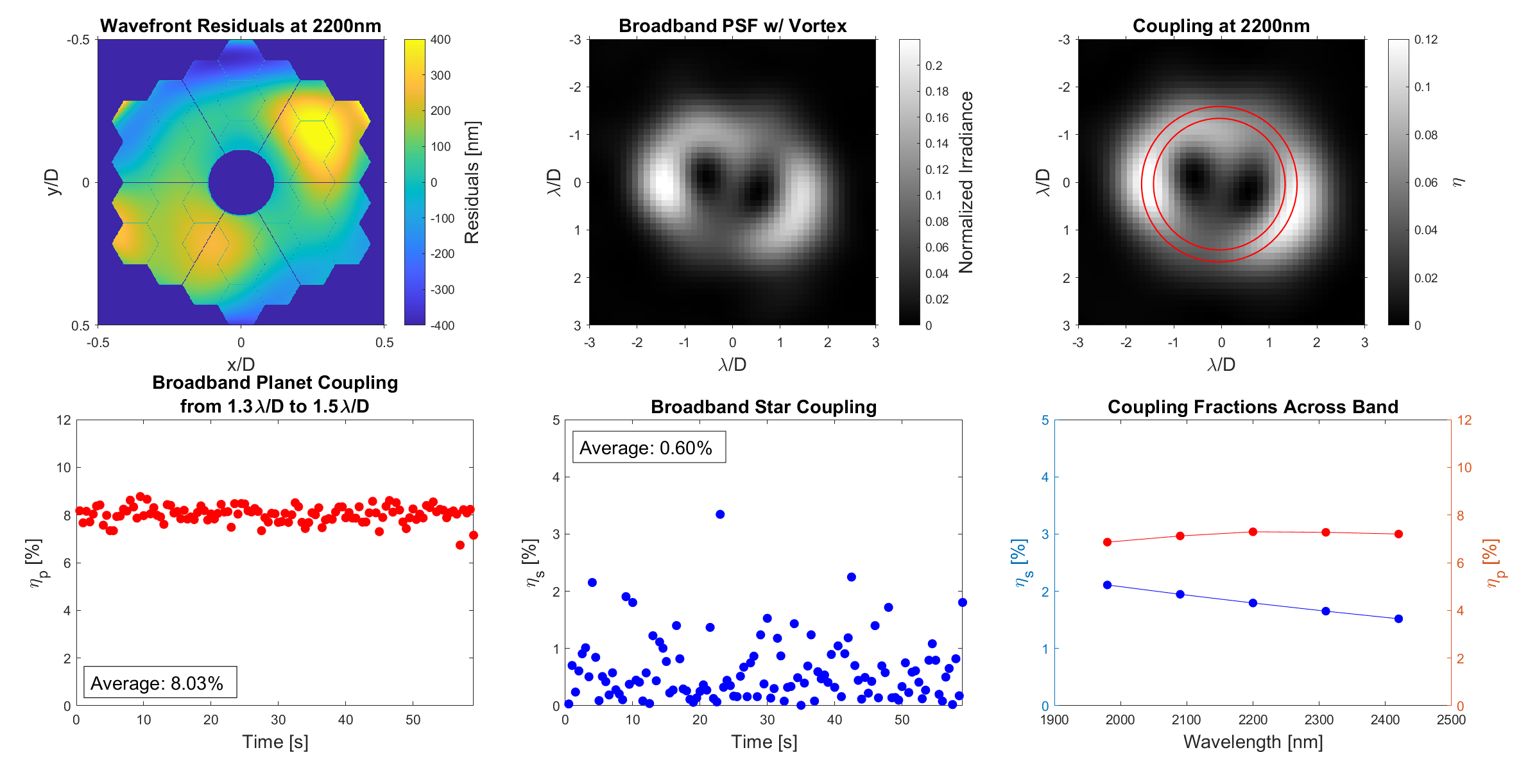}
    \label{fig:Charge2Simulation}
\end{subfigure}
\begin{subfigure}{\textwidth}
    \centering
    \vspace{-5mm}
    \caption{Potential charge 1 KPIC VFN case (Online Video 2).}
    \includegraphics[width=\linewidth]{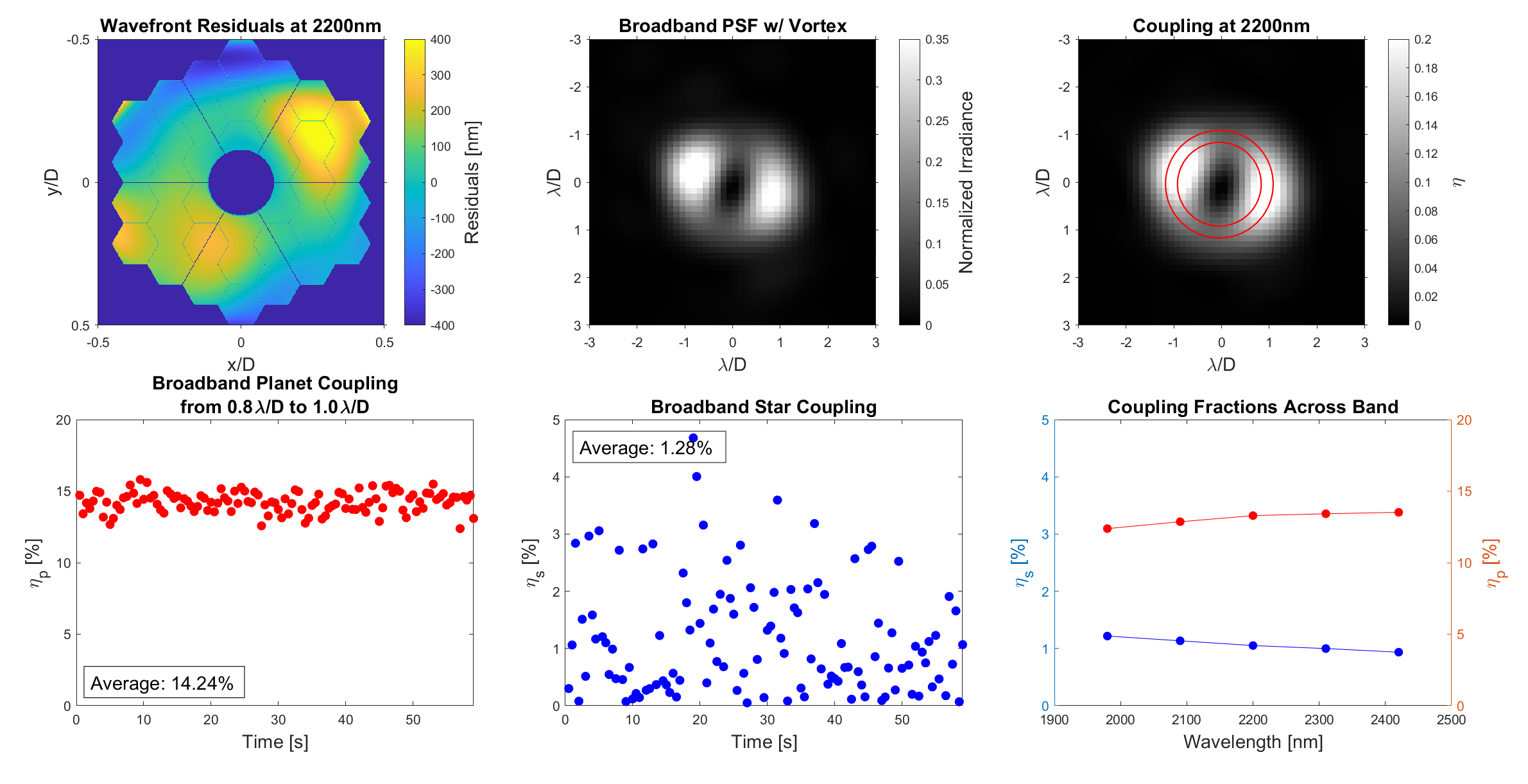}
    \label{fig:Charge1Simulation}
\end{subfigure}
\caption{(Video online) KPIC VFN performance simulations given the on-sky performance of the PyWFS as well as the predicted ADC residuals with an ideal vortex mask of (a)~charge 2 and (b)~charge 1. The upper left plot in each is the reconstructed PyWFS residuals. The upper middle plot is the system PSF while the upper right is the corresponding coupling efficiency for all points within a $6\times6\lambda/D$ field of view. The lower left plot shows the average coupling for a planet within the region bound by the two red circles in the coupling map. The lower middle plot is the star coupling. The lower right plot is the planet and star coupling at each of the 5 sample wavelengths across the $K$-band.}
\label{fig:SimulatorResults}
\end{figure}

In the simulator, we decompose the real wavefront data into Zernike coefficients and then reconstruct the wavefront as projected onto the real Keck pupil. Figure~\ref{fig:SimulatorResults} shows the final frame, or time-step, of the simulator for a charge 2 (Fig.~\ref{fig:Charge2Simulation}) and charge 1 (Fig.~\ref{fig:Charge1Simulation}) VFN case. As such, the reconstructed wavefront at this final time step is shown in the upper left plots. We then add in the tip/tilt residuals from Fig.~\ref{fig:RawPyWFSData}b at the given time sample as well as the predicted chromatic dispersion left over from the ADC to get the net pupil phase as a function of wavelength. We apply the vortex phase assuming an ideal, achromatic charge 1 or charge 2 vortex mask and calculate the resulting PSF at five sample wavelengths across the band. The upper middle plot of Figs.~\ref{fig:SimulatorResults}a,b shows the broadband PSF, which would be imaged on the tracking camera. 

To get the coupling efficiencies across the band, we compute Eq.~\ref{eqn:couplingeff} for every point in the field at each sample wavelength. The resulting 2D coupling map is shown in the upper right plot of Figs.~\ref{fig:SimulatorResults}a,b for the central operating wavelength of $2.2\mu m$. We calculate the predicted planet coupling, $\eta_p$, as the average for all points between $0.8\text{-}1.0\lambda/D$ for the charge 1 case and $1.3\text{-}1.5\lambda/D$ for charge 2 to account for uncertainties in the planet location. The region used in this average is shown between the two red circles superimposed on the 2D coupling map and encompasses the separation at which the theoretical peak coupling occurs for each vortex charge. The resulting planet coupling, plotted against time, is shown in the lower left plot of Figs.~\ref{fig:SimulatorResults}a,b. The time-averaged planet coupling is shown at the bottom left of this plot. Under these assumptions, we find that the predicted time-averaged planet coupling is 8\% for a charge 2 vortex and just over 14\% for charge 1.

In order to compute the star coupling, $\eta_s$, we must choose where to place the SMF in our simulations. Due to the tip/tilt residuals, the PSF moves around with respect to the SMF much faster than we can track and compensate for with the current TTM control loop. This means that taking the null point in the coupling map at each frame would be an unfair representation of the actual on-sky performance since we would be assuming that we can align the PSF with the SMF core infinitely fast. We therefore take the average of all the coupling maps and find the optimal null location in this time-averaged map. We then place our simulated fiber at that location and compute the coupling efficiency there at each time sample. This is representative of what we expect from a realistic TTM control loop which will tend to average out the tip/tilt residuals. The resulting star coupling is shown in the lower middle plot of Figs.~\ref{fig:SimulatorResults}a,b. The time-averaged null depth is reported in the upper left corner of this plot. Given the wavefront residuals used in this simulation as well as the predicted ADC residuals, we get an average null depth of 0.6\% $(6\times10^{-3})$ for the charge 2 case and 1.3\% $(1.3\times10^{-2})$ for charge 1. The final, lower right, plot in Figs.~\ref{fig:SimulatorResults}a,b shows the instantaneous coupling efficiency for the star and planet at each of the 5 sample wavelengths across our $K$-band simulation. This assumes a flat spectrum for both the star and planet. 

Thus, Fig.~\ref{fig:Charge2Simulation} represents the predicted performance for the planned KPIC Charge 2 VFN mode while Fig.~\ref{fig:Charge1Simulation} shows a possible charge 1 case for comparison. As expected, the planet coupling at the peak planet location increases to 14\% with the charge 1 vortex mask but the null depth also degrades to 1.3\%. The tradeoff is whether the decrease in null depth is worth the access to closer companions.
%Therefore, given the current KPIC AO performance, the charge 1 case provides access to a closer region around the star as well as improved planet coupling at the cost of a marginal decrease in null depth.
%Given that the integration time, $\tau$, required to detect a planet in the photon-limited regime scales as $(\tau\sim\eta_s/\eta_{p}^2)$\cite{Ruane2018_VFN}, the degradation in null depth is close to compensated for by the improved planet coupling but we gain access to a region closer to the star. 

The images shown in Fig.~\ref{fig:SimulatorResults} are stills of the final frame in the simulation. The video version of these figures, showing the instantaneous wavefront, PSF, and coupling, is available in the online copy of these proceedings. 

The results of these simulations are promising and indicate that the current PyWFS performance is sufficient for obtaining $<10^{-2}$ nulls while coupling 8\% of the planet light with a charge 2 vortex as planned. We can expect that this performance will improve further once the high-order DM is integrated into the KPIC system. However, these simulations are preliminary and there are other effects that may impact the VFN performance including realistic polarization aberrations, on-sky ADC residuals, and non-common path aberrations. %Furthermore, these simulations only account for throughput losses due to the nulling effect of VFN, they do not account for the other losses accrued prior to the SMF face.
We are also working towards simulating the characterization capabilities of the KPIC VFN mode by injecting simulated planet atmospheric spectra, accounting for the planet-star contrast ratios, applying the throughput losses in the rest of the system, and attempting to extract molecules from the resulting signal\cite{Wang2017}.

\clearpage

\section{Summary and Future Outlook}
\label{sec:Conclusion}

VFN is a promising new concept that will soon be tested on-sky with the KPIC instrument. The concept has been demonstrated in the lab with a visible wavelength, monochromatic null depth of $6\times10^{-5}$ and planet coupling of 16\% with a charge 1 vortex. We have also demonstrated a polychromatic null of $4.2\times10^{-4}$ and planet coupling of $4.5\%$ in 10\% bandwidth visible light with a charge 2 vortex mask. We have laid out a path for upgrading the Caltech VFN testbed in order to further improve the VFN performance as well as move to longer wavelength demonstrations in preparation for the deployment of a KPIC VFN mode. Using measurements taken from the KPIC pyramid wavefront sensor during an on-sky engineering night, we have simulated the predicted performance of the VFN mode at first light. 

Moving forward, we plan to continue developing the VFN concept on the upgraded testbed at Caltech. Some of the planned experiments include switching to longer wavelengths and larger bandwidths as well as testing scalar vortex designs. We also plan on installing a low-order DM into the VFN testbed to start developing wavefront control algorithms that are tailored to match VFN requirements. 

For on-sky operations of a VFN mode on KPIC, we will need to develop automated algorithms and procedures for finding and maintaining a deep null. These are already being developed on the VFN testbed and will eventually be moved to the telescope system along with the wavefront control algorithms. Once the VFN architecture is integrated, we will start commissioning the VFN mode following a similar fashion to what was done for the direct planet spectroscopy mode. Thus, we will target bright binaries first to confirm that we can null one and extract the spectral signatures of the other on the NIRSPEC spectrograph. As the acquisition, tracking, and extraction algorithms improve, we will move towards known, RV-detected, young giant planets and eventually transition to blind detection surveys around promising M-dwarf stars.

VFN may enable us to collect and analyze the reflected spectra of exoplanets that are inaccessible to conventional coronagraphic techniques; i.e. at separations $<1\lambda/D$. These are mainly young giant planets of which there is a large population detected by RV and transit techniques. Thus, a VFN mode on a large ground-based telescope (e.g. Keck or TMT) or on a space-based telescope (e.g. the LUVOIR mission concept\cite{Bolcar2018_LUVOIR}) would be complementary to proposed coronagraphic instruments and would increase the scientific yield with minimal modifications to the system.

\acknowledgments  
Part of this work was carried out at the Jet Propulsion Laboratory, California Institute of Technology, under contract with the National Aeronautics and Space Administration (NASA).

\clearpage

%%%%%%%%%%%%%%%%%%%%%%%%%%%%%%%%%%%%%%%%%%%%%%%%%%%%%%%%%%%%%
%%%%% References %%%%%
\small
\bibliography{Library}   %>>>> bibliography data in report.bib
\bibliographystyle{spiebib}   %>>>> makes bibtex use spiebib.bst

\end{document}